# Spintronics and Pseudospintronics in Graphene and Topological Insulators

Dmytro Pesin and Allan H. MacDonald

Department of Physics, University of Texas at Austin, Austin TX 78712-1081, USA

ABSTRACT

**The two-dimensional electron systems in graphene and in topological insulators are described by massless Dirac equations. Although the two systems have similar Hamiltonians, they are polar opposites in terms of spin-orbit coupling strength. We briefly review the status of efforts to achieve long spin relaxation times in graphene with its weak spin-orbit coupling, and to achieve large current-induced spin polarizations in topological-insulator surface states that have strong spin-orbit coupling. We also comment on differences between the magnetic responses and dilute-moment coupling properties of the two systems, and on the pseudospin analog of giant magnetoresistance in bilayer graphene.**

*I: Introduction* --- The central goals of spintronics [1] are to understand mechanisms by which it is possible to achieve efficient electrical control of spin-currents and spin-configurations, and to discover materials in which these mechanisms are prominently exhibited. Because of the obvious relationship to magnetic information storage technologies, the possibility of applications is always in the background of spintronics research topics, and sometimes jumps to the foreground to spectacular effect. Nevertheless, the problems that arise in this field are often intriguing from a fundamental point of view, and many topics are pursued for their intrinsic interest. Many of the most active themes of spintronics research are reviewed elsewhere in this issue. This Progress Article is motivated by recent interest in two new types of electronically two-dimensional (2D) material: graphene layer systems [2][3], and surface-state systems of 3D topological insulators (materials that act as bulk insulators but have topologically protected surface states) [4][5]. We briefly review recent work that has explored these materials from the spintronics [6] point-of-view, provide a perspective of how graphene and topological insulator systems fit into the broader spintronics context, and speculate on directions for future research.

Studies of graphene-based 2D electron systems (2DESs) and of two and three dimensional topological insulators are among the most interesting and active current topics in materials physics. The two systems have closely related, although still distinct, electronic properties. We restrict our attention primarily to 3D topological insulators in which the low-energy degrees of freedom are surface-state electrons that are described by 2D Dirac equations. The same equations describe low-energy $\pi$-band electrons that are confined to a single graphene sheet. Below we refer to the two systems generically as 2D Dirac systems (2DDSs).

The spintronics field can be organized as summarized in Fig. 1. The most important distinction is between conductors with magnetic order and conductors without magnetic order. The 2DDSs are in the second category so far, although one can imagine hybrid systems in which spins behave collectively because of proximity effects or because of deliberately introduced dilute moments[7][8][9][10].

Spintronics exists as a topic largely because of the difference (two or more order of magnitude) between electron velocities in conductors and the speed of light, which almost always makes spin-orbit coupling weak. Spins are therefore usually almost conserved: that is, their relaxation times are normally much longer than other characteristic electronic time scales. In ordered systems, weak spin-orbit coupling leads to magnetic anisotropy energies that determine the energetically favorable magnetization orientations in a crystalline lattice. The anisotropy energies are extremely small compared to magnetic condensation energies, which measure how much the energy is lowered by ordering magnetically. For this reason, the spin-densities that can be induced by transport currents [11] are sufficient to induce magnetization switching, even though they are small in a relative sense. On the other hand, the changes in electrical transport properties that can accompany switching are often large, because they reflect the full strength of exchange interactions in the magnetic state. (Thermal transport is also sensitive to magnetic configurations [12].)

Paramagnetic conductors are not spin-polarized in their ground state, but because of spin-orbit coupling transport currents (which break time-reversal invariance) can induce spin-densities. Moreover currents can be weakly spin-polarized even in non-magnetic materials [13]. In spintronics, paramagnetic conductors with especially weak spin-orbit coupling are desirable as 'spin-conservers' that can transmit spin-encoded information across a device with high fidelity. On the other hand paramagnetic conductors with strong spin-orbit coupling are desirable because they can be 'spin-generators' when combined with transport currents. Generally speaking, the surface states of topological insulators have exceptionally strong spin-orbit coupling and could provide an interesting example of a spin-generator system, whereas graphene $\pi$-bands have weak spin-orbit coupling and have potential as a spin-conserver system. At the time of writing, fewer potential applications have been identified for the more subtle spintronics effects that occur in non-magnetic conductors, like the materials that we will discuss here, although recent work on magnetization reversal induced by the spin Hall effect near heterojunctions between magnetic and non-magnetic materials[14][15] suggests that their potential has not yet been fully realized. In our view the most intriguing possibilities for 2DDSs are gate-tunable dilute-moment or nanoparticle-array magnetism at topological-insulator or graphene surfaces, and layer-pseudospin giant magnetoresistance in graphene bilayers.

At the present time, experimental work on graphene-sheet 2DESs is much more advanced than work on topological-insulator surface states. The study of topological insulators has so far advanced slowly because of materials problems, which include the lack of real insulating behavior in many cases. Although these materials issues [16] are critically important, they are not addressed in this Progress Article.

**Spintronics in Graphene Sheets**

The states near the Fermi level of a graphene sheet are $\pi$ electrons. In neutral graphene, the $\pi$-bands are half-filled, and the Fermi level lies at the energy of the Brillouin-zone corner where there is a linear band crossing between $\pi$ conduction and valence bands - that is, at the Dirac point. For energies close to the Dirac point, the band Hamiltonian $H$ is accurately approximated by a two-dimensional Dirac equation:

$$H = \hbar v(\tau_z \sigma_x k_x + \sigma_y k_y) \quad (1)$$

where $(k_x, k_y)$ is 2D momentum measured from the band-crossing point, $v \approx 10^8$ cm s$^{-1}$ is the Bloch state velocity at the Dirac point, $\tau_z$ is a Pauli matrix that acts on the valley degree of freedom that distinguishes the two inequivalent band crossing points, and $\sigma_{x,y}$ are Pauli matrices that act on the honeycomb lattice's sublattice degree-of-freedom. It is common in the graphene literature to view the sublattice degree-of-freedom as a pseudospin, with the eigenstates of $\sigma_z$ localized on A or B sublattices (Fig.2.) Using this language, the Hamiltonian is purely off-diagonal in pseudospin; it represents hopping between sublattices that vanishes at the Dirac point because of destructive interference between the three nearest-neighbor hopping paths. Graphene sheets are ambipolar, - that is, their Fermi energies can be shifted by approximately $\pm 0.3$ eV relative to the Dirac point by gating. Surface states of topological insulators, discussed in the next section, have similar 2D Dirac bands as explained in Fig. 2.

The spin-conserving potential of graphene is due to the weakness of spin-orbit interactions at energies close to the Dirac point of an intrinsic inversion symmetric sheet. These interactions take the form [17][18][19]

$$H_{SO}^{G,I} = \Delta_I \sigma_z \tau_z s_z \quad (2)$$

where $s_z$ is a Pauli matrix that acts on the spin degrees-of-freedom. The form of the interaction is determined entirely by symmetry and corresponds to a staggered potential which has opposite signs on opposite sublattices (the $\sigma_z$ factor), and for each sublattice opposite signs for opposite spins (the $s_z$ factor). The valley dependent $\tau_z$ factor indicates that this potential has opposite signs when its dependence on triangular lattice position is Fourier-transformed at the two inequivalent Brillouin-zone-corner Dirac points $K$ and $K'$. The magnitude of $\Delta_I$ is smaller than the characteristic spin-orbit scale of $2p$ electrons in carbon $\xi \sim 6$ meV, because [18][19] spin-orbit interactions vanish when projected onto the atomic $\pi$ bands of a flat sheet. Spin-orbit coupling in a flat graphene sheet is weaker than in the curved graphene sheets of nanotubes [18][20][21]. Using a simple tight-binding approach to describe spin-orbit induced mixing between $\pi$- and $\sigma$- bands leads to the estimate $\Delta_I \sim 1\mu$ eV. *Ab initio* calculations [22][23] predict values that range from $\sim 1\mu$ eV to $\sim 50\mu$ eV. Although the small value of this coupling constant adds to the difficulty of estimating its value accurately, there is no doubt that it is small compared to characteristic spin-orbit energy scales. Because of its small value, it has also not yet been possible to directly measure $\Delta_I$ in graphene. Experiment has, however, placed an upper bound of $\sim 100\mu$ eV on the value of $\Delta_I$ (ref. [24]).

The spin relaxation associated with the intrinsic spin-orbit interaction occurs via the Elliot-Yafet disorder-scattering spin-relaxation mechanism [25][26]. Because band eigenstates do not have a pure spin, spin-reversals can be induced even by a spin-independent disorder potential. In the graphene case, however, the component of spin perpendicular to the surface is not relaxed ($\tau_\perp \to \infty$) because the intrinsic spin-orbit interaction commutes with $s_z$. For the in-plane spin polarization, the spin relaxation rate for electrons with the Fermi energy, $\varepsilon_F$, is [25][26]

$$\frac{1}{\tau_{\parallel}} \approx \frac{\Delta_I^2}{\varepsilon_F^2} \frac{1}{\tau} \qquad (3)$$

where $\varepsilon_F$ is measured from the band-crossing point. Note that when the Elliot-Yafet mechanism applies, the spin relaxation time, $\tau_{\parallel}$, is proportional to the momentum relaxation time, $\tau$; this is the identifying property of the Elliot-Yafet relaxation mechanism. The latter property is quite natural, because in the Elliot-Yafet mechanism spin relaxes only during collisions and the spin- and momentum-relaxation rates must therefore be proportional.

To obtain a conservative theoretical estimate of the Elliot-Yafet spin-flip time, we adopt the above experimental limit of $\sim 100 \mu$eV as an estimate for the intrinsic spin-orbit strength, and assume a mobility of 3,000 cm$^2$V$^{-1}$s$^{-1}$ at the carrier densities of $\sim 10^{12}$ cm$^{-2}$ used in most experimental spin-relaxation studies. The resulting spin-relaxation time is $\sim 50$ ns. If we instead assumed a value of $\Delta_I$ in the mid-range of theoretical estimates ($\sim 10 \mu$eV), the spin-relaxation time would exceed a microsecond and the corresponding spin-diffusion length $\lambda_{sd} = \sqrt{D\tau_{\parallel}}$ would be $\sim 300 \mu$m. Here the diffusion constant $D$ is given by $v^2\tau/2$. If $\Delta_I$ was closer to the tight-binding estimates, $\tau_{\parallel}$ and $\lambda_{sd}$ would be even longer.

Unfortunately, this simple and happy scenario has not yet been realized; observed spin memory times and lengths are still much shorter than these values. Early Hanle spin-precession studies [27][28][29][30][31][32][33] found that the spin-diffusion length $\lambda_{sd}$ was indeed proportional to the charge scattering time $\tau$, (Fig. 3) but the measured carrier spin lifetimes in graphene sheets were in the $100-200$ ps range. Furthermore, the sense of the anisotropy between the in-plane and out-of-plane relaxation rates observed experimentally [34] was that $\tau_{\perp} < \tau_{\parallel}$, opposite to expectations based on Elliot-Yafet relaxation due to intrinsic spin-orbit coupling. These observations indicated that although spins were indeed relaxed by the Elliot-Yafet mechanism, non-intrinsic spin-orbit interaction mechanisms must play an essential role.

Motivated by these observations, a variety of different extrinsic spin-relaxation mechanisms were studied theoretically [35][36][37][38]. Recently it was demonstrated [39] that relaxation induced by ferromagnetic contacts can play a role in limiting the apparent spin lifetime in Hanle experiments performed with transparent contacts. Because this spin-relaxation mechanism can be mitigated by fabricating pinhole-free tunnel barriers between the graphene sheet and spin-injectors and detectors, the discovery is an encouraging one. At present, the largest spin relaxation rates that have been achieved in single layer graphene [40][41] are ~0.5 ns. But because samples prepared in similar ways can show very different spin-relaxation times, the source of the spin-flip scattering that limits the spin-relaxation time in a particular material is usually not known. Longer spin-lifetimes, breaking the 1-ns barrier, have been achieved in bilayer graphene, possibly owing to enhanced screening or reduced surface exposure in this carbon allotrope [40][41] as summarized in Fig. 3. Importantly, it was observed that in bilayer graphene a regime can be achieved in which the spin relaxation time is inversely proportional to the momentum relaxation time. As we now explain, this observation signals the presence of spin-split bands due to broken inversion symmetry.

Quite generally, any material with broken inversion symmetry can exhibit the Dyakonov-Perel spin-relaxation mechanism [42]. This mechanism accounts for changes in spin direction between collisions with impurities. When Bloch states are spin-split, propagating electrons precess under the influence of a momentum-dependent effective magnetic field. Dyakonov-Perel relaxation can occur in single and bilayer graphene because of Rashba coupling due to gating electric fields or ripples. Near the Dirac point, the spin-orbit coupling Hamiltonian for the Rashba interaction is [17]

$$H_{SO}^{G,R} = \Delta_R (\sigma_x \tau_z s_y - \sigma_y s_x) \qquad (4)$$

where $\Delta_R$ is the strength of Rashba spin-orbit coupling. If the effect of ripples is taken into account, $\Delta_R$ acquires a random position-dependent contribution [36]. The dependence on momentum direction enters through the Dirac-Hamiltonian (1) via its dependence on the in-plane components of the lattice pseudospin vector $(\sigma_x, \sigma_y)$. It should be noted that the above Hamiltonian will contribute to Elliot-Yafet-type relaxation in addition to the DP relaxation discussed below. However, the Elliot-Yafet rate is always smaller than the Dyakonov-Perel rate contribution in good conductors, that is, when $\varepsilon_F \tau \gg 1$, which normally holds in graphene except at small carrier densities.

The Dyakonov-Perel relaxation rate can be estimated using simple physical arguments. For weak spin-orbit coupling ($\Delta_R \ll \hbar/\tau$), the Bloch state lifetime $\tau$ is not long enough to resolve the small spin-orbit splitting. Spins then relax via a process of random precessional walks on the spin Bloch sphere with step length $\Delta_R \tau / \hbar$ and step time $\tau$. A spin with a non-equilibrium orientation will relax after a time $\tau_{DP}$ such that

$$\frac{\Delta_R \tau}{\hbar} \sqrt{\frac{\tau_{DP}}{\tau}} \sim 1. \qquad (5)$$

These considerations imply that

$$\tau_{DP} \sim \frac{\hbar^2}{\Delta_R^2 \tau}, \qquad (6)$$

and that the spin-diffusio length is

$$\lambda_s = \sqrt{D \tau_{DP}} \sim \frac{\hbar v}{\Delta_R}. \qquad (7)$$

For single-layer graphene, the Rashba spi-orbit strength is estimated to be tens of microelectronvolts per 1V nm$^{-1}$ of external electric field. For a density of $10^{-12}$ cm$^{-2}$ we find that $\Delta_R$ ~0.5 $\mu$eV and conclude that gate-induced Rashba spin-orbit coupling is not strong enough to be responsible for the spin-lifetimes observed in single-layer graphene. The random Rashba spin-orbit field due to graphene ripples can be stronger, however[18][36], and still stronger effects are possible in bilayer graphene.

Because graphene has an exposed 2D surface, it turns out that it can be used not only as a spin-conserver but also as a spin-generator. A very large increase in Rashba spin-orbit coupling strength can be achieved by covalent bonding with absorbates[35][43][44][45][46][47][48]. These findings suggest the possibility of creating lithographically defined spintronics devices based on graphene with a designed network including both spin-conserving and spin-generating regions. Spin-generation effects, including the spin Hall effect and current-induced spin-densities, are discussed at greater length below in the context of topological insulators, which always have strong spin-orbit interactions as we shall emphasize.

**Topological Insulator Surface-State Spintronics**

Strong three-dimensional topological insulators [4][5][49] possess an odd number of 2D Dirac bands localized on the bulk material's surface. The simplest case, that of a single Dirac cone, is realized in the $Bi_2Se_3$ family of topological insulators. Because of its role in the early experimental exploration of topological insulators, $Bi_2Se_3$ has been referred to as the hydrogen atom of topological insulators[4]. The surface-state 2DDSs of topological insulators are similar to those present in graphene, but differ in three essential respects as indicated in Fig.2. First of all, for topological insulators there is only one conduction- and valence-band surface state at each 2D momentum, compared with the four states (two for spin times two for valley) present in the graphene case. Although fundamental, this distinction is often academic because spin and valley coupling effects in graphene are too weak to play a significant role in most observable properties. Second, the position of the Fermi level relative to the Dirac point in these materials is determined by bulk physics, not surface physics; the Fermi level does not generically lie at the Dirac point [50] when the TI is neutral. Pristine neutral graphene is, in contrast, guaranteed to have a Fermi level that intersects the Dirac point. This distinction is also largely academic in practice, as the position of the Fermi level at the surface of a topological insulator can be adjusted with external gate voltages, just as it is in most graphene studies. The third distinction is the most important one, particularly for spintronics. The 2D massless Dirac equation for TI surface states couples spin, not sublattice-pseudospin, to momentum. The potential role of topological insulator surface states in spintronics is that of a spin-generator, not a spin-conserver, as we discuss below.

The simplest approximate Hamiltonian for a surface state of a strong topological insulator, including both orbital and Zeeman coupling to an external magnetic field is

$$H_{TI} = v\vec{\sigma}\cdot\vec{z}\times(\vec{p}-\frac{e}{c}\vec{A}_{ext}) - g\mu_B\vec{\sigma}\cdot\vec{b}_{ext} \qquad (8)$$

In the above Hamiltonian, $\sigma$ is a Pauli matrix that acts on spin (not sublattice pseudospin) degrees of freedom, $\vec{z}$ is chosen to be the direction normal to the surface of a TI, $\vec{p}$ is the canonical momentum, $g$ is the Lande g-factor for the surface electrons (assumed isotropic for simplicity), $\mu_B$ is the electron Bohr magneton, and $\vec{A}_{ext}$ and $\vec{b}_{ext}$ are the external vector potential and magnetic field. When the topological insulator surface is proximity coupled to a ferromagnetically ordered material, $\vec{b}_{ext}$ can include an exchange-field contribution. The orbital term in the Hamiltonian (equation (8)) is identical to the Rashba spin-orbit coupling term in the Hamiltonian of a two-dimensional electron gas [1], but here it is the entire Hamiltonian. Spin-orbit interactions are as strong as they could possibly be on any topological-insulator surface. This fact explains why the surface of a topological insulator cannot be used as a spin-conserver. Such a surface is almost always in the regime in which the spin-orbit field is larger than the elastic scattering rate. The spin-relaxation time is therefore always of the same order as the transport scattering time, which depends on sample quality but is typically much shorter than $10^{-9}$ s. At topological-insulator surfaces there is no motivation for identifying the spin-relaxation time as a separate time scale.

To compare the spin-generator potential of a topological-insulator surface to other 2DEGs, we take the ratio between the current-induced spin density and the total electron density as a figure of merit. From the Hamiltonian (equation (8)) it follows that the electric current operator is in fact proportional to the spin density operator. (The same property applies in the graphene case when spin is substituted by pseudospin.) The explicit expression is

$$\vec{j} = ev\vec{\sigma} \times \vec{z} \qquad (9)$$

Using the Drude formula, $\sigma_{Drude} = ne^2\tau/m$ with effective mass $m = p_F/v$, for the conductivity of a 2DDS on a topological insulator surface leads to the following estimate for the spin polarization induced by a transport current [51]

$$\frac{s_y}{n} = \frac{j}{evn} = \frac{eE\tau}{p_F}, \qquad (10)$$

where $E$ is the applied electric field, and $p_F$ is the Fermi momentum of surface electrons. The ratio of current-induced spin-density to the full electron density is equal to the ratio of the transport drift velocity to the Fermi velocity. The corresponding figure of merit for, say, Rashba-coupled 2DEG is smaller by an additional factor of $\Delta_R/E_F \ll 1$ (ref. [52]). In this sense a topological-insulator surface is a much more effective spin generator than a regular 2DEG. Larger spin polarizations can be achieved in ferro/normal hybrid systems, but topological-insulator surfaces offer the possibility of gate voltage control.

Another way to generate spin polarization in a spin-orbit coupled system is via the spin Hall effect (SHE) [53] in which transverse spin currents appear in response to an applied electric field. These spin currents can then lead to spin accumulation at the boundaries of the system or in adjacent conductors. In general, one expects that both intrinsic mechanisms [54] related to the spin-orbit coupling in band structure, and helical scattering effects [55] due to spin-orbit impurity potential scattering can contribute to the SHE. For a disordered topological-insulator surface, however, there are two reasons not to expect a large spin Hall signal in spite of the strong spin-orbit interactions. First, it is known that for a disordered 2D system with Rashba-like spin-orbit coupling that is linear in momentum the d.c. intrinsic SHE vanishes [56][57][58][59][60]. Second, even if cubic in momentum corrections to Hamiltonian (equation (8)), or extrinsic contribution to the SHE are considered, the spin relaxation length coincides with the transport mean free path on a topological-insulator surface, and thus the spin accumulation is restricted to boundary regions with a mean-free-path size. This makes the observation of the resultant spin accumulation a challenge, although not an impossibility (see for example ref. [61]). These considerations motivated studies of SHE on topological-insulator surfaces to focus on mesoscopic samples in the ballistic regime [62][63]. It was found that the resultant spin polarizations exceed their semiconductor counterparts by at least an order of magnitude. It is important to note that even though the Dirac electrons on the two opposite surfaces of a topological insulator have opposite chiralities, and thus could partially cancel current-induced spin densities and spin currents, their spatial separation, in principle, makes it possible to address each surface separately. For instance, a gap could be induced on the side surfaces of the sample, at least in principle, by coating with a magnetic film. (See below on why a magnetic field is able to induce a gap on a surface of a topological insulator.) The experimental observation of these effects awaits the discovery of techniques that can be used to prepare topological-insulator samples with surfaces of the required quality.

Although the dissipationless voltage-induced spin Hall current of topological-insulator surface states is likely small, the correlation between velocity and spin directions implies that large (non-dissipative) spin-currents flow in equilibrium.  These spin currents likely have observable consequences only if they can flow from the TI surface to another material with magnetic order or weaker spin-orbit coupling.  It seems clear that equilibrium spin currents do not satisfy this condition.  The correlation between velocity and spin directions will also tend to suppress large-angle scattering, as does the analogous property of graphene sheets, and therefore to support large longitudinal conductivities.  These properties of the surface states of 3D topological insulators faintly echo those of the helical liquid 1D edge state systems associated with two-dimensional topological insulators [64][65][66]. In the case of the latter edge state, the only possible type of elastic scattering is backscattering between states that are spin and orbital time-reversed partners, which is absent in the absence of time-reversal-symmetry breaking because of Kramer's theorem.

We now turn to the effects of external perturbations that break time-reversal symmetry. Because the presence of gapless surface states in strong and weak topological insulators is protected by time-reversal invariance, the surface-state system is sensitive to perturbations that break this symmetry. In the Hamiltonian (equation (8)), the external magnetic (or exchange) field is the source of time-reversal invariance breaking. Because of the strictly linear dispersion in the Hamiltonian (equation (8)), an in-plane magnetic field enters the Hamiltonian in precisely the same way as a static, spatially-constant vector-potential. There is therefore no paramagnetic response on a TI surface to uniform static in-plane exchange or magnetic fields. This means that for strictly linear dispersion there is also no transport-current-induced spin polarization perpendicular to the plane. Perpendicular-to-plane spin-polarizations in systems with in-plane spin-orbit effective magnetic fields cannot in any case be explained solely on the basis of Boltzmann transport theory. They were nevertheless discovered [67] experimentally in semiconductor 2DEGs and later explained [68] theoretically in terms of a delicate balance between collision and generation terms in a quantum-kinetic transport theory.

Magnetic fields that are applied in the $\vec{z}$ direction, on the other hand, do produce large effects. An exchange field applied in the $\vec{z}$ direction (see below) opens up a gap at the Dirac point and induces [4][5][69][70] a half-quantized quantum Hall effect when the Fermi level lies in this gap. Non-zero charge anomalous Hall effects [71][72] are present at all carrier densities. An external magnetic field in the $\vec{z}$ direction opens up a large gap at the Dirac point and smaller gaps at many other filling factors with half-odd-integer quantized Hall conductivities. The Hall conductivity increases as the Fermi level moves further away from the Dirac point, while the corresponding gaps get smaller. A number of interesting magnetoelectric [49][73][74][75], and magneto-optical phenomena [70][76][77][78][79][80][81][82][83]  are associated with weak time-reversal-symmetry breaking at the surface of a topological insulator.

The properties of topological-insulator surface states in a perpendicular external magnetic field are very similar to those of graphene sheets because the orbital response to a magnetic field, which tends to dominate, is identical in the two cases. The response of such surface states to exchange coupling with  magnetically ordered systems is, however, entirely different from that of a graphene systems. Two ideas have been suggested for engineering a system in which this exchange coupling is present: i) introducing magnetic impurities [84] that can order due to surface-state mediated interactions, and ii) proximity coupling to evaporated magnetic films [85].

The former possibility was considered theoretically in refs. [9] and [10]. Spin-momentum locking at the surface leads to an unconventional RKKY interaction. [9][10][86][87] which gives rises to a rich phase diagram [9][10] that is briefly summarized in Fig. 4. The maximum temperature for magnetic impurity ferromagnetism was estimated to be ~30K, in decent agreement with an experimental observation of 13K [84]. The relatively low transition temperature means that this type of order does not have practical utility in $Bi_2Se_3$ systems, although the same mechanism could lead to much higher transition temperatures in as yet undiscovered topological insulators. A more practical approach in the near term might stem from the demonstration [85] that magnetic proximity effects due to an iron coating make the surface of a strong topological insulator ferromagnetic at room temperature.

As a medium for local-moment coupling, graphene sheets might have some advantages over topological-insulator surface states because their spin-susceptibility is electrically tunable and can be made large by shifting the Fermi level far from the Dirac point. Local moments can be induced in graphene by defects including vacancies and hydrogen or fluorine adatoms [8]. Unfortunately, the coupling between these defects and graphene's $\pi$ orbitals seems so far to be weak [88][89].

**Pseudospintronics**

Any two-component quantum degree of freedom that electrons possess in addition to their orbital degrees of freedom can behave in a manner that is mathematically equivalent to spin, and can therefore be viewed as a pseudospin. The analogy between spins and pseudospins is useful only in special circumstances since pseudospins lifetimes are normally short, and the identified degrees of freedom are often not continuous across sample boundaries. As mentioned earlier, graphene's sublattice degree of freedom can be viewed as a pseudospin and enters the continuum model Dirac equation in precisely the same manner that real spin enters the Dirac equation for thr surface states of topological insulators. In the pseudospin language, the Hamiltonian consists only of a pseudospin-orbit coupling term with an effective magnetic field that is linear in momentum and points in the same direction as momentum. Much of what we have explained about the role of spin in topological-insulator surface states, applies equally well to pseudospin in single-layer graphene. Just as in the spin case, we can expect that charge currents in single-layer graphene will be accompanied by pseudospin currents. It is, however, not obvious how these pseudospin currents could be measured.

Another example of a two-valued degree of freedom that is often usefully viewed as a pseudospin is the layer degree of freedom in a bilayer electron system. In the case of bilayer graphene there are four carbon atoms per unit cell, but it turns out [90] that there is also, at least at energies below the interlayer bonding energy $\gamma_1 \sim 0.3\,\mathrm{eV}$, a spin-1/2 pseudospin degree of freedom which can be interpreted as labeling a layer rather than a sublattice. The pseudospin effective magnetic field in the balanced bilayer band structure is proportional to momentum squared in magnitude and has an x-y plane orientation angle that is twice the orientation angle of the momentum. Because the $z$ component of pseudospin in the bilayer case corresponds to layer-polarization, an electric field applied between layers acts like a $z$-direction pseudospin field. This pseudospin field can readily be tuned experimentally from being near zero to being the largest energy scale in the system. No similar experimental control is available for the sublattice pseudospin of single-layer graphene. Unlike the case of a ferromagnetic metal in which external magnetic fields have a negligible influence on the overall density of states, the pseudospin field in bilayer graphene can open up a sizable gap - effectively turning a metal into a semiconductor. This surprising feature of the electronic properties of bilayer graphene seems to offer the best potential [91] for digital electronics based on graphene.

In 3D systems, there are two types of spin Hall response associated with a current in the $x$ direction: i) currents of $y$ spins in the $z$ direction and ii) currents of $z$ spins in the $y$ direction. For a finite-thickness quasi-2D system, currents in the vertical $z$ direction should lead to spin accumulations of opposite sign near the top and bottom surfaces. In bilayer graphene we can therefore anticipate that a current in the $x$ direction will lead to a $z$ direction pseudospin current that flows in the $y$ direction, and to $y$ direction pseudospin accumulations that have opposite signs in the top and bottom layers. These effects have so far not been extensively explored.

The large and technologically useful magnetotransport properties of metallic ferromagnets occur even though the coupling between individual charge carriers and a magnetic field is rather weak. Because the collective magnetic orientation coordinate of a ferromagnet balances external fields and current effects against only weak anisotropy field effects, mechanisms that would be weak for individual electrons are strong collectively and sufficient to achieve complete reversal of the magnetization. When the magnetization direction is reversed in a part of a magnetic circuit, the effective potential seen by an electron of a given spin is changed by a large amount - in fact by the size of the exchange spin-splitting in that particular magnet, which can be of the order of electronvolts. We do not expect collective pseudospin order to be present at room temperature in graphene systems, although recent evidence [92][93] does suggest that bilayer graphene has collective pseudospin order at low temperatures when the carrier density is very close to the neutrality point. Bilayer graphene does, however, have the unique feature that there is a readily available experimental capability which can vary the $z$ component of the pseudospin field over wide ranges, easily larger than the carrier Fermi energy. This property can be used, for example, to devise [94][95] pseudospin analogs of ferromagnetic metal spin-valve devices, Fig. 5.

**Looking forward**

Graphene and topological-insulator surface states have very similar Hamiltonians, but represent two distinct extremes of paramagnetic conductor spintronics. The Dirac equation in graphene captures strong coupling between momentum and the sublattice pseudospin degree of freedom. The coupling between graphene's orbital and spin degrees, on the other hand, is so weak that it is normally neglected in analyses of graphene electronic properties and, in fact, has not yet been measured directly. Spin-orbit coupling is weak in graphene not only because carbon is a relatively light element, but also because of its planar geometry reduces the coupling between the Fermi-level $\pi$ electrons and $\sigma$ electrons. This means that the spin-creator spintronic effects, transport-current induced spin-densities and spin-polarized currents, are extremely weak in graphene. Graphene's potential in spintronics is as an unusually effective spin-conserver, which can transmit spin-information over extremely large distances. Spin-lifetimes in initial experiments were limited by hybridization with contact electrodes. At present, spin-lifetimes in graphene are still limited by extrinsic disorder effects that have not been definitively identified. Given the rapid pace of recent progress, we can expect that future efforts will achieve spin-lifetimes that are limited only by intrinsic effects. These experiments should enable a definitive measurement of the intrinsic spin-orbit coupling strength at the Dirac point, whose value still remains quite uncertain. This development would be satisfying, as it might finally allow the quantum spin Hall effect to be observed in the material for which it was originally proposed [17]. An exception to the weak spin-orbit coupling properties of graphene arises when it is hybridized with certain metals.

Topological insulator surface states are strongly spin-orbit coupled, unlike the electron states in graphene sheets. Spin lifetimes will inevitably be close to Bloch-state lifetimes. The interesting spintronic properties of such surface states are characterized by their spin Hall and current-induced spin-density properties. The experimental and theoretical exploration of these effects is still at an early stage, but we can expect that the second effect will be stronger than the first. The ratio of current-induced spin densities to the overall charge density is usually small by a factor of drift velocity divided by Fermi velocity and also by a factor of spin-orbit coupling energy divided by Fermi energy. In the case of topological-insulator surface states, the second small factor is absent. Spin Hall effects, on the other hand, are known to be weak in systems in which the spin-orbit coupling term is approximately linear in momentum. This condition applies to states near the Dirac point, suggesting that spin Hall effects will be weak in spite of the very strong spin-orbit interactions.

Because of their electrical tunability, 2DESs are attractive as magnetic stiffness agents in systems containing dilute magnetic moments. In the case of (Ga,Mn)As, the Mn local moments induce too much disorder when inserted into a quantum well and robust 2D ferromagnetism has been elusive. It is interesting to speculate on the potential of 2DDSs in graphene and topological insulators as hosts for diluted-moment ferromagnetism. It is certainly clear that the two materials will behave very differently. In mean-field theories, which are accurate when the moment density is comparable to the carrier density and carriers are able to provide adequate magnetic stiffness, the magnetic transition temperature is proportional to the carrier spin susceptibility. In graphene, the susceptibility is nearly perfectly isotropic because of weak spin-orbit interactions. Attractively, the susceptibility is also strongly sensitive to carrier-density, vanishing with the carrier density-of-states at the Dirac point. In the case of topological-insulator surface states, the susceptibility is strongly anisotropic, large only for out-of-plane fields, and reaches a maximum when the Fermi level is placed at the Dirac point. Any magnetic state based on coupling between magnetic moments that is mediated by topological-insulator surface states will have the attractive feature of strong perpendicular anisotropy.

Graphene and multilayer graphene $\pi$-electron systems possess sublattice degrees of freedom at each crystal momentum, which may be viewed as pseudospins. In both single-layer and bilayer graphene, low-energy states have two-component pseudospins that are mathematically equivalent to the electronic spin. The band Hamiltonians in both cases have very strong pseudospin-orbit coupling that is similar to the strong spin-orbit coupling of topological-insulator surface states. Because the sublattice pseudospin is equivalent to layer in the case of bilayer graphene, electric fields perpendicular to the layer couple much more strongly to pseudospin than practical magnetic fields are able to couple to real spins.

The unique electronic structure features shared by 2DDSs in graphene and topological-insulator surface state are likely to provide them with a number of roles in spintronics, and the rapid progress over the past few years, partially chronicled here, points to future surprises. The most intriguing opportunities, in our view, are for gate tunable magnetism in graphene sheets and at topological-insulator surfaces (perhaps magnetism involving coupling between nanoparticle arrays) and electrically tunable pseudospin properties in multilayer graphene systems that can be used to create pseudospin analogues of giant magnetoresistance and other established spintronics effects.

**Acknowledgements**

We acknowledge financial support from the US Army Research Office (ARO) under award number MURI W911NF-08-1-0364. We thank D. Abanin, I. Grigorieva, J. Sinova, A. Veligura, and B. van Wees for discussions.

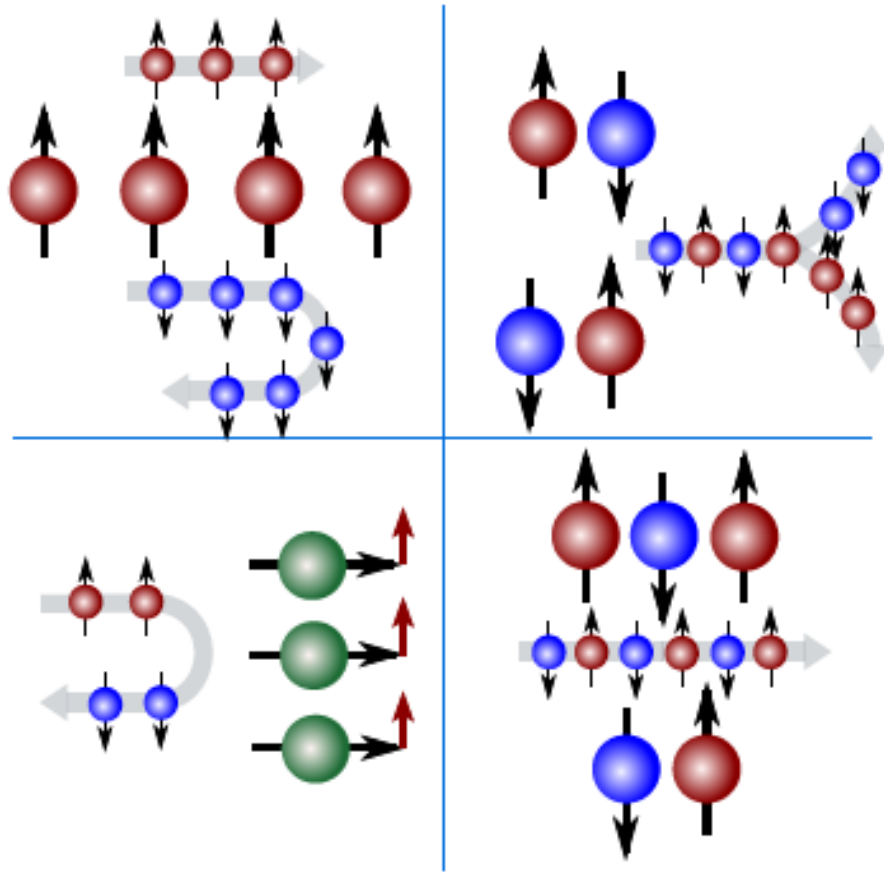

**Figure 1 | Overview of spintronics.** In all panels, the large spins represent overall magnetization and the small spins represent the transport electrons. The field of spintronics is divided in the first place between the study of magnetically ordered conductors (left panels) and the study of paramagnetic metals or semiconductors (right panels). Within each class, one can study effects in which electric fields alter spin configurations (bottom panels) and complementary effects (top panels) in which effective magnetic fields due to spin-orbit, exchange, or magnetostatic interactions influence transport properties. The four panels in this figure (anticlockwise from the top left) schematically illustrate i) giant magnetoresistance in which variation in magnetization direction increases backscattering and hence resistance; ii) Andreev reflection of spins in non-collinear magnetic systems that leads to spin-transfer torques and current-induced spin-reversal; iii) current-induced spin polarization in paramagnetic conductors; and iv) the spin Hall effect and spin currents in paramagnetic conductors. Spin-transport effects in paramagnetic conductors always require spin-orbit interactions. This Progress Article concentrates on right-panel phenomena in topological insulators and in monolayer graphene, and left-panel phenomena in bilayer graphene.

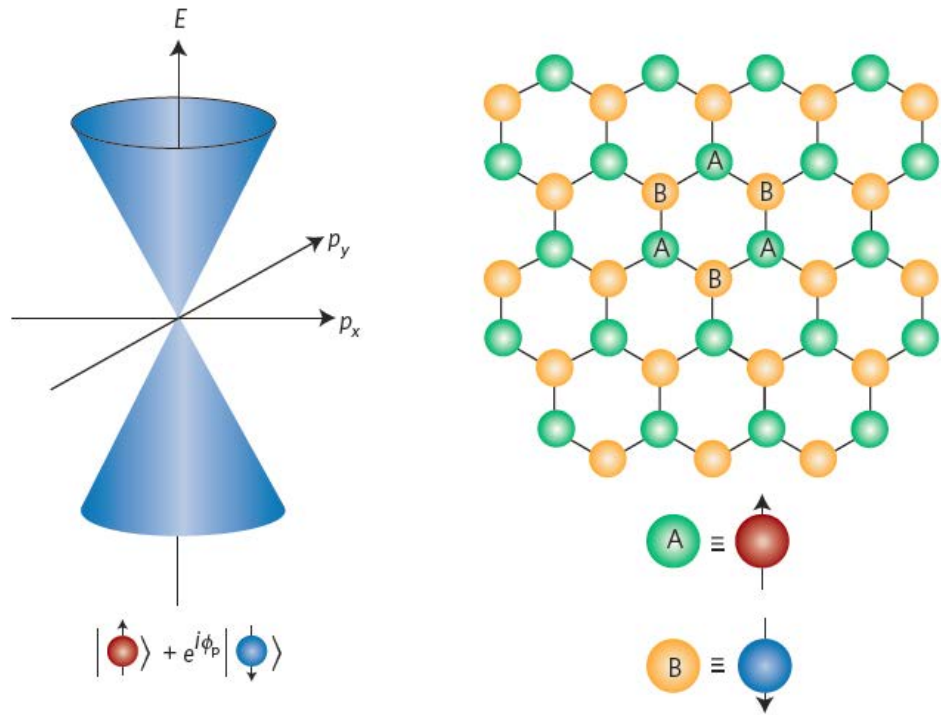

**Figure 2 | Dirac cones in graphene and in topological insulators.** Topological-insulator surface states and $\pi$-band states in graphene (left panel) are described by a 2D Dirac equation with strong coupling between momentum and spin in the topological-insulator case and between momentum and sublattice pseudospin in the graphene case. Topological-insulator surface states are non-degenerate and are coherent equal-weight linear combinations of two spin components with a momentum-dependent phase difference, $\varphi_p$. The pseudospin representation of the honeycomb sublattice degree of freedom is illustrated in the right panel. The $\pi$-band eigenstates in graphene are coherent equal-weight linear combinations of their two honeycomb sublattice components, denoted by A and B, and have spin- and momentum-space valley degeneracies.

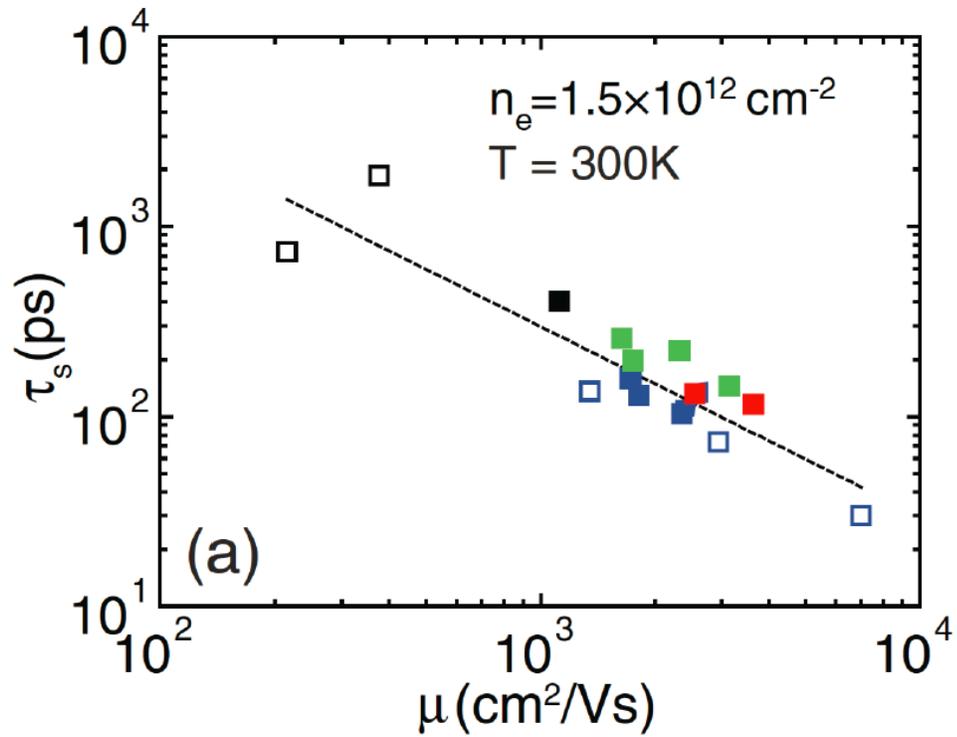

**Figure 3 | Hanle effect measurements of room-temperature spin-relaxation times in a variety of bilayer graphene samples [41].** Mobilities, $\mu$, of the samples vary by two orders of magnitude. Similar results were reported in ref. [40]. The property that the spin-relaxation time decreases with increasing mobility suggests that the Dyakonov-Perel spin-relaxation mechanism is operative in these systems. Reprinted with permission from ref. [41], © 2011 APS.

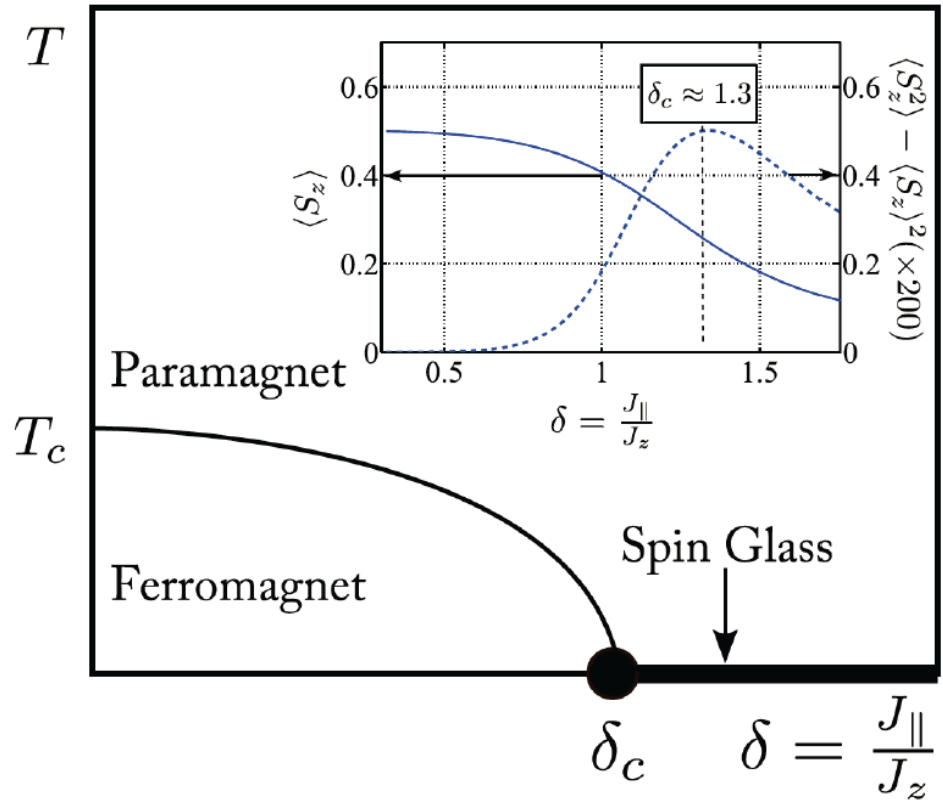

**Figure 4 | Phase diagram for magnetic adatom magnetism mediated by topological insulator surface states.** Exchange coupling constants $J_{\parallel,\perp}$ characterize the coupling between in-plane and out-of-plane components of the spins of the magnetic adatom and topological-insulator surface electrons. Inset: dependence of spin-1/2 adatom disorder-averaged magnetization (solid line) and magnetization fluctuations (dashed line) on the exchange anisotropy, $\delta = J_{\parallel}/J_{\perp}$ (solid line). The position of the quantum phase transition between the out-of-plane ferromagnet and spin glass phases, $\delta_c \approx 1.3$, is inferred from the finite-size condition that magnetization has decreased by 50% from its value at $J_{\parallel} = 0$. The position of the transition is further confirmed by a maximum in the magnetization fluctuations at $\delta = \delta_c$. Reprinted with permission from ref. [9], © 2011 APS.

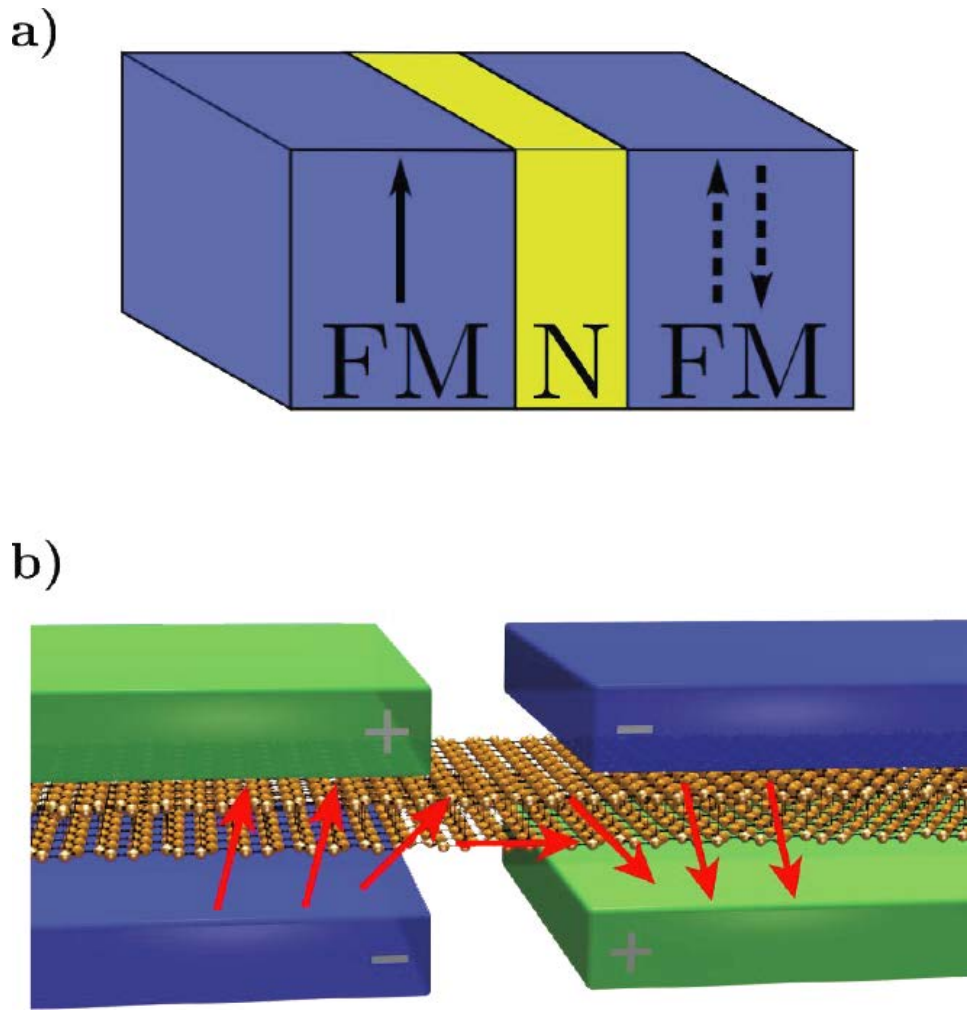

**Figure 5 | Comparison between a regular ferromagnetic metal spin valve device and a bilayer graphene pseudospin valve device. a.** In the case of the usual spin valve, the strong dependence of resistance on the relative orientation of two ferromagnetic elements placed in series in a magnetic circuit is due to the strongly spin-dependent transport of ferromagnets and to interface scattering from the exchange potential change at the interface. (See also the upper left panel of Fig. 1). **b.** In a bilayer graphene pseudospin valve, the relative pseudospin alignment of bilayer segments placed in series is controlled by external gates. The increase in device resistance when the gate voltages have opposite sign is due entirely to scattering of injected pseudospin-polarized electrons off the pseudospin texture (indicated by red arrows) created by the gate voltages. Part b reprinted with permission from ref. [94], © 2009 APS.